\documentclass[a4paper,fleqn]{cas-dc}
\renewcommand{\printorcid}{} 
\usepackage[numbers]{natbib}

\usepackage{lineno,hyperref}
\usepackage{rotating, graphicx,picinpar}
\usepackage{multirow}
\usepackage{array}
\usepackage{color}
\usepackage{ulem}
\usepackage{lipsum}
\usepackage{graphicx}
\usepackage{upgreek}
\usepackage{multicol,multirow}
\usepackage{amsmath,amssymb,amsfonts}
\usepackage{mathrsfs}
\usepackage{amsthm}
\usepackage{textcomp}
\usepackage{xcolor}

\usepackage{caption}
\usepackage{subcaption}
\usepackage{paralist}
\usepackage{multirow}
\usepackage{algorithm}
\usepackage{algorithmic}
\usepackage{amsmath}
\usepackage{multicol}
\usepackage{tabularx}
\usepackage{rotating}
\usepackage{mathrsfs}
\def\tsc#1{\csdef{#1}{\textsc{\lowercase{#1}}\xspace}}
\tsc{WGM}
\tsc{QE}
\tsc{EP}
\tsc{PMS}
\tsc{BEC}
\tsc{DE}
\begin{document}
\let\WriteBookmarks\relax
\def\floatpagepagefraction{1}
\def\textpagefraction{.001}
\shorttitle{Discovery of mixing characteristics for enhancing coiled reactor performance through a Bayesian Optimisation-CFD approach}
% \shorttitle{Data-driven approach to identify flow features enhancing plug flow performance in coiled tube reactors }
\shortauthors{Basha et al.}

%\title [mode = title]{Data-driven discovery of mixing characteristics for enhancing plug flow performance}            
\title [mode= title] {Discovery of mixing characteristics for enhancing coiled reactor performance through a Bayesian Optimisation-CFD approach}
%\tnotemark[1,2]

%\tnotetext[1]{This document is the results of the research
% project funded by the National Science Foundation.}

%\tnotetext[2]{The second title footnote which is a longer text matter
 %  to fill through the whole text width and overflow into
  % another line in the footnotes area of the first page.}

\author[1]{Nausheen Basha}[type=editor,
  %                      auid=000,bioid=1,
 %                       prefix=Sir,
 %                      role=Researcher
]
%\cormark[1]
\ead{nausheen.basha@imperial.ac.uk}
%\ead[url]{www.cvr.cc, cvr@sayahna.org}

%\credit{Conceptualization of this study, Methodology, Software}
\author[1]{Thomas Savage}
\author[2]{Jonathan McDonough}
\author[1]{Ehecatl Antonio Del-Rio Chanona}
\author[1]{Omar K. Matar}

%\ead{cvr3@sayahna.org}
%\ead[URL]{www.sayahna.org}

%\credit{Data curation, Writing - Original draft preparation}

\address[1]{Department of Chemical Engineering, Imperial College London, South Kensington Campus, London SW7 2AZ, United Kingdom}
\address[2]{School of Engineering, Newcastle University, Merz Court, Newcastle upon Tyne NE1 7RU, United Kingdom}
%\cortext[cor2]{Principal corresponding author}
%\fntext[fn1]{This is the first author footnote. but is common to third
 % author as well.}
%\fntext[fn2]{Another author footnote, this is a very long footnote and
 % it should be a really long footnote. But this footnote is not yet
  %sufficiently long enough to make two lines of footnote text.}

%\nonumnote{This note has no numbers. In this work we demonstrate $a_b$
 % the formation Y\_1 of a new type of polariton on the interface
 % between a cuprous oxide slab and a polystyrene micro-sphere placed
 % on the slab.
  %}

\begin{abstract}
Processes involving the manufacture of fine/bulk chemicals, pharmaceuticals, biofuels, and waste treatment require plug flow characteristics to minimise their energy consumption and costs, and maximise product quality. 
One such versatile flow chemistry platform is the coiled tube reactor subjected to oscillatory motion, producing excellent plug flow qualities equivalent to well-mixed tanks-in-series '$N$'.
In this study, we discover the critical features of these flows that result in high plug flow performance using a data-driven approach. 
This is done by integrating Bayesian optimisation, a surrogate model approach, with Computational fluid dynamics 
that we treat as a black-box function to explore the parameter space of the operating conditions, oscillation amplitude and frequency, and net flow rate. 
Here, we correlate the flow characteristics as a function of the dimensionless Strouhal, oscillatory Dean, and Reynolds numbers to the reactor plug flow performance value ‘$N$’. 
% Optimal performance for a simulated coil occurs at $St= 1.9$ and $De_0/Re= 2.4$ after evaluating just 36 iterations, with a performance of $N= 29.4$
Under conditions of optimal performance (specific examples are provided herein), the oscillatory flow is just sufficient to limit axial dispersion through flow reversal and redirection, and to promote Dean vortices. 
This automated, open-source, integrated method can be easily adapted to identify the flow characteristics that produce an optimised performance for other chemical reactors and processes. 
\end{abstract}

%%%\begin{graphicalabstract}
%%%\includegraphics{figs/grabs.pdf}
%%%\end{graphicalabstract}

%%%\begin{highlights}
%%%\item Research highlights item 1
%%%\item Research highlights item 2
%%%\item Research highlights item 3
%%%\end{highlights}


\maketitle
\begin{sloppypar}
\section{Introduction \label{introduction}} 

\noindent The industrial sector in 2021 was responsible for emitting 9.5Gt of CO\textsubscript{2}, accounting for a quarter of the emissions worldwide \cite{iea2021industry}. With mixing processes being one of the key features of most modern industrial processes, they are expected to consume large amounts of energy and thereby release sizeable amounts of emissions. 
Additionally, poor mixing commonly leads to significant economic losses from a combination of lower yield (\$100 million), complications in the scale-up and process development (\$500 million), and further lost opportunities from unsuccessful products that never reach the market \cite{Paul2004}. 

Plug flows are a mixing pattern where the fluid elements undergo enhanced radial mixing and negligible axial mixing. 
They have a wide range of applications, including manufacturing pharmaceuticals \cite{Porta2016}, fine/bulk chemicals \cite{Loponov2017}, biodiesel production \cite{Tabatabaei2019}, treatment of agro-industrial waste \cite{Eftaxias2021}, and wastewater treatment \cite{Pan2016}.
This type of mixing pattern promotes faster reactions, good heat and mass transfer rates, better process control, and reduction of undesired by-products \cite{Coker2001}. 
Coiled tube reactors have been shown to achieve excellent plug flow characteristics, and are simple to fabricate, operate and maintain. 
Therefore, these reactors are in demand for various process intensification \cite{Soni2019}, and other applications such as manufacturing \cite{Wu2017}, polymerisation \cite{Parida2014}, heat exchange \cite{Kumar2006}, food processing \cite{Singh2013}, biotechnology \cite{Klutz2015} and even as cardiovascular stents \cite{Cookson2019}. 
Even modest improvements in plug flow performance will therefore have a wide impact on product yield, sustainability, capital, energy and operational costs, justifying further research in this area. 

Fluid flowing through curved tubes experiences a centrifugal force that leads to strong secondary flow structures in the form of twin counter-rotating vortices known as Dean vortices \cite{Dean1927}. 
These vortices are responsible for promoting radial mixing whilst reducing axial dispersion. The formation and distortion of the Dean vortices with the variation of Reynolds number have been widely studied experimentally for coiled tubes \cite{Liou1992,Kovats2020}, spiral \cite{Nivedita2017} and serpentine channels \cite{Reddy2020} under non-oscillating or steady inlet conditions. 
The development of secondary flows in coiled tubes is only observed when $Re>40$ \cite{Mansour2017}, but radial mixing in comparison to a straight tube is not observed until $Re>300$. 
This has the potential to limit the operational window. 
For example, reactions requiring longer residence times may necessitate impractically long channels. 
One way to decouple the throughput from the mixing characteristics is to superimpose oscillations onto the net flow. 
Mixing can then be tuned using the oscillatory conditions whilst the residence time can be tuned using the flow rate \cite{McDonough2019}. 
The oscillatory flow introduces complex time-dependent flow structures that affect mixing performance.

Decoupling of plug flow/mixing from the net flow has also been widely reported with the Oscillatory Baffled Reactor (OBR) platform, where oscillatory flows affect axial and radial mixing through the formation of vortices behind obstructions such as orifice baffles \cite{Howes1991,Mackley1991,Roberts1995}. 
Oscillatory flows can be characterised by the Strouhal number ($St$) and oscillatory Reynolds number ($Re_o$). 
These describe the vortex propagation lengths and vortex size, respectively. 
A wide range of time-dependent flow phenomena have been observed for a range of $Re_o$ and $St$.  
For large values of $St$, the flow field is dominated by viscosity, limiting the development of vortices due to strong dissipation of the flow energy \cite{Sobey1983}. 
With decreasing $St$ and increasing $Re_o$, time-periodic vortex patterns are formed, which develop in strength with further decrease in $St$ and increase in $Re_o$, eventually resulting in three-dimensional, unsteady, aperiodic flows \cite{Howes1988,Nishimura1991, Roberts1996, Nishimura1989}. 
OBRs that use helical spring baffles rather than orifice baffles produce further complexities, where the flow switches between radially dominated and swirl-dominated under varying oscillation conditions \cite{Phan2012,Mcdonough2017,McDonough2019}. 
A similar analogy can be expected for coiled tube reactors where oscillation intensity related to $Re_o$ and $St$ could produce flow patterns that involve vortex shedding and swirling that, in turn, might impact plug flow performance in unpredictable ways. 
This idea is further supported by McDonough et al. \cite{McDonough2019a}, where the plug flow was observed to switch ‘on’ and ‘off’ with changes in oscillation intensity. 
Optimal plug flow conditions were reported at $De_o/Re= 2-8$ and $St= 1-2$, but the underpinning flow patterns in the coil affecting the performance were not explored. 

Given the importance of plug flow, the versatility of the helical coil reactor, and the advantages of coiled reactors under oscillatory flow conditions, as summarised in the foregoing, in this study we seek to identify the flow characterisitcs that correspond to the optimal plug flow region and also explore the transition of the flows into these optimal regions from non-optimal regions as the oscillation intensity changes.
The use of dimensionless numbers ($Re$, $Re_o$, $St$, $De_o$) has helped in the design and development of high-efficiency reactors \cite{Bellhouse1973,Ni1995,Cox2022,Abbott2013,Avila2020}.
Computational Fluid Dynamics (CFD), CFD-led explorations have been extensively carried out for improving the design of reactors, where design improvements can be identified based on visualisations of the flows \cite{bao2013optimal,ding2010cfd,hapke2021optimization}.
However, these are expensive and are often based on human intuition of the parameter space, which makes this approach ineffective at balancing the number of evaluations needed to identify truly optimal solutions. 
Therefore, in this study, we will combine CFD with data-driven methods in a sample-efficient manner to discover the flow patterns that are responsible for optimal plug flow performance. 
This will involve exploring the complex parameter space of oscillatory conditions.

CFD-based data-driven experimental design and optimisation of reactors has been carried out using evolutionary algorithms such as NSGA-II \cite{Deb2002,Chen2016,Mansour2020} or MOGA-II \cite{Poloni2000,uebel2016cfd}. 
However, these require a significant number of expensive evaluations to be effective. 
These CFD simulations can be replaced with cheap surrogate models; relevant applications of evolutionary algorithms with surrogate models have been reported \cite{Ong2005,Roßger2018,Rigoni2010} though it should be noted that 
these approaches still rely on randomness in both exploration and exploitation during evaluations which need rigorous searching methods based on probability. 
Bayesian optimisation (BO) has been put forward as a strategy to circumvent the limitations of traditional (costly) exploration of design spaces. 
BO has been shown to be capable of finding optimal (global) solutions with a minimal number of function evaluations \cite{van2022data} through
a flexible surrogate model (typically a Gaussian Process \cite{Seeger2004}) to stochastically approximate the (generally) expensive objective function. 
This cheap surrogate model is sequentially updated with new information on the design space based on the values of the acquisition function. 
This acquisition function also chooses the next point for evaluation based on a certain metric or `policy' with an end-goal that accelerates the iterative design process, generally balancing exploration and exploitation. 
It is well-suited to cases where the evaluations or black-box functions (CFD simulations in this case) are expensive \cite{Jones1998,Chen2017,Li2017,Park2015,Lei2021}. 
Therefore, in recent years, there has been a growing number of applications for BO with CFD \cite{Diessner2022,Daniels2022,Renganathan2021,Morita2022}. 
However, less than a handful are focused on chemical reactors \cite{Park2018,Kim2022}, including the ones from the present authors \cite{savage2022deep,savagedeep}, let alone the exploratory use of BO for parameter design maps and flow pattern investigation of chemical reactors. 

In this study, we will combine BO with CFD to explore and exploit the parameter space of frequency and amplitude (oscillatory conditions) for a fixed low $Re$ for a coiled tube reactor geometry. 
Through exploration of the design space, an ensemble of parameters resulting in high plug performance is obtained and categorised into optimal or non-optimal regions. 
With exploitation, parameters resulting in the most-optimal plug flow performance are achieved. 
We then evaluate the flow characteristics pertaining to various levels of plug flow performance, enabling us to `discover' the underpinning desirable features. 
The main challenges addressed by this study are \textit{i}) to apply a sample-efficient methodology for CFD-enabled data-driven exploration of parameter spaces to optimise plug flow performance; \textit{ii}) to represent the explored parameter space with mixing flow characteristics in order to gain deeper insights into performance which can guide future reactor development; and \textit{iii}) to uncover specific mixing flow characteristics for an optimal condition. 
We expect that the adopted methodology and insights gained from this work are transferable to the design of other efficient chemical reactors and related devices.

 % It also seems likely that nature may have exploited some of the phenomena described in this paper. The pulsed nature of blood flow may ensure that stagnant zones behind “natural baffles” are avoided, or might interact harmonically with some natural instability resulting in improved transfer properties. 
%%%%%%%%%%%%%%%%%%%%%%%%%%%%%%%%%%%%%%%%%%%%%%%%%%%%%%%%%%
\section{Methodology \label{section2}} 
\noindent In this section, we discuss the optimisation and design space identification procedure in detail. The optimisation of the plug flow reactor is described mathematically and the principles of Bayesian optimisation are set out with a specific focus on the exploration term to explore the parameter space. The details related to the computational model and flow solver are then explained alongside their integration with the optimisation procedure.  Finally, we introduce the key parameters that play a crucial role in characterising the mixing flow.

\subsection{Problem formulation}\label{problemsection}
\begin{figure}
    \centering
    \includegraphics[width=\columnwidth]{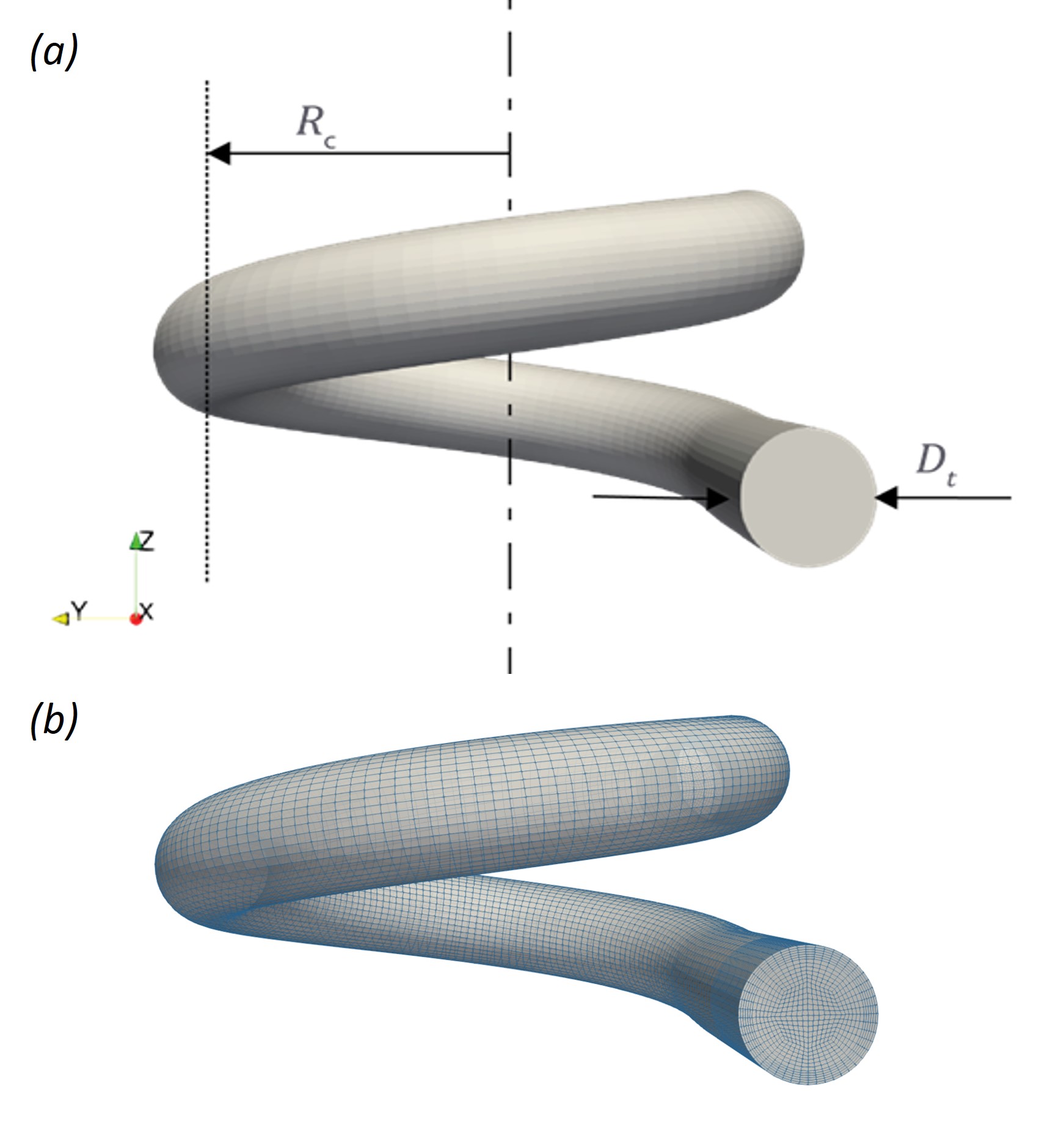}
    \caption{(a) CAD model of the single turn coiled tube labelled with the geometric parameters, and (b)  Computational grid for the coiled tube with the domain extents.}
    \label{fig:meshed_image}
\end{figure}

\noindent We aim to maximise the plug flow performance $N$ for a typical coiled reactor. The reactor considered in this study has a single turn with $R_c= 12.5 \ mm$ and $D_t= 5 \ mm$ (Figure \ref{fig:meshed_image} a). 
% We deliberately chose a single coil to reduce computational time as the performance obtained from a single turn can be related to a coil with multiple turns.
The fluid flowing through the coil has density $\rho$ and viscosity $\mu$ and the flow is assumed to be incompressible. 
The inlet flow speed $v_{\rm in}$ is given by
\begin{equation}
    v_{\rm in} = v_o + v_{\rm steady},
\end{equation}
where $v_{\rm steady}$ is the speed associated with the steady part of the inlet velocity, and $v_o$ characterises the oscillatory flow imposed at the inlet:
\begin{equation}
    v_o=2 \pi f x_o \sin(2\pi f t);
\end{equation}
here, $x_o$, $f$, and $t$ represent the oscillation amplitude and frequency, and time, respectively. 
The Reynolds number is defined as 
\begin{align}
    Re=\frac{\rho v_{\rm steady} D_t}{\mu},
\end{align}
%$Re=\rho v_{\rm steady} D_t/\mu$, 
and the steady flow considered in the present work corresponds to $Re=50$. The aim of imposing the oscillatory flow component at the inlet is to achieve plug flow despite this relatively low $Re$ value. 
%Superposition of an oscillatory velocity $v_o$ on the steady inlet velocity $v_{\rm steady}$ 
%is needed to achieve plug flows at the flow inlet velocities considered int he present work %at this low $Re=\rho v_{\rm steady} D_t/\mu$, 
%and this is expressed through two parameters: amplitude $x_o$ and frequency $f$, giving an oscillatory velocity of \textit{$v_o$= 2$\pi$f$x_0$sin(2$\pi$ft) }. The total inlet velocity, ($v_{in}$) is then, \textit{$v_{in}$= $v_o$ \ + $v_{steady}$ }. $v_{steady}$ corresponds to a set Reynolds number, $Re = 50$.

\begin{figure}
    \centering
        \includegraphics[width=\columnwidth]{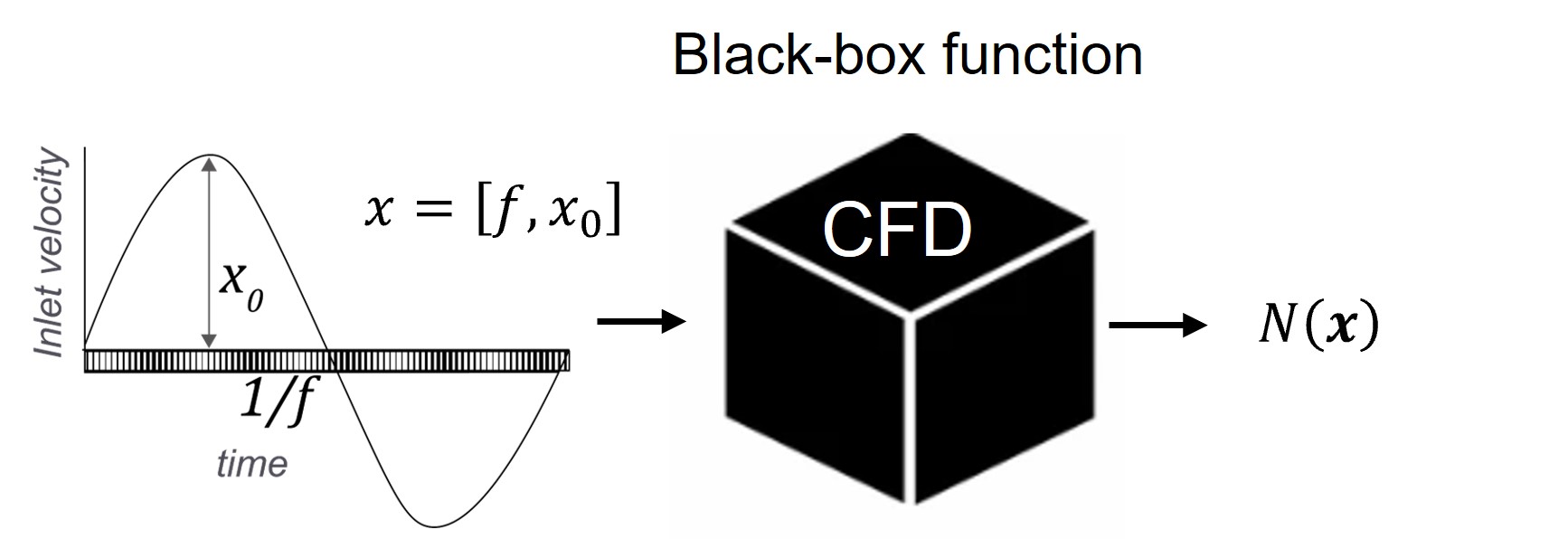}
    \caption{Graphical illustration of problem formulation.}
    \label{fig:black-box}
\end{figure}

The optimisation problem can then be formulated as 
\begin{align}
    \mathbf{x}^* = \mathop{\mathrm{argmax}}_{\mathbf{x}\in \mathcal{X}}\; f(\mathbf{x}). \label{black_box_problem}
\end{align}
where the black-box function, or objective function $f(x)= N(x)$, is evaluated with the decision variables ${x= [x_0,f]}$, and bounded in the set of possible operating conditions $\mathcal{X}$, as illustrated in Figure \ref{fig:black-box}. The limits are $x_o \in [{1-8}]$ mm and $f \in [{1-8}]$ Hz, set according to operational practicalities. 

To ensure wider applicability of the problem formulation, we introduce dimensionless parameters based on the decision variables, namely $Re_o$, $St$, and $De_o$, which are defined as follows:
\begin{align}
    Re_o=\frac{2 \pi f x_o \rho D_t}{\mu},
\end{align}

\begin{equation}
    St = \frac{R_c}{2 \pi x_o}, ~~~~~~
    De_o = Re_o\sqrt{\frac{D_t}{2R_c}}.
\end{equation}

%%%%%%%%%%%%%%%%%%%%%%%%%%%%%%%%%%%%%%%%%%%%%%%%%%%%%%%%%%
\subsection{Bayesian Optimisation}\label{BO}

\noindent The purpose of this article is %to discover the underlying characteristics of the behaviour of pulsed-flow within the coiled tubes. A key aspect of this investigation is to not only determine the general flow characteristics, but 
to identify the optimal characteristics of oscillatory flows in coiled tubes. To achieve this, there is a need to explore the parameter space effectively while keeping the focus on the optimal regions. There are two main approaches that can be employed to perform this exploration: Design of Experiments (DoE), and optimisation-based techniques. The DoE approach provides a systematic way to explore the parameter space by selecting points based on a predetermined sampling strategy. Alternatively, optimisation-based techniques search for the optimal solution by iteratively refining the search space based on the objective function. In this study, we choose to apply Bayesian Optimisation (BO) as our primary method of exploration. BO is a global, derivative-free optimisation technique that builds a probabilistic model of the objective function and uses it to select the most promising points to evaluate. The acquisition function, which guides the selection of points, balances exploration and exploitation based on a trade-off parameter, denoted as $\kappa$.

The BO acquisition function, $\alpha$, is formulated as follows:
\begin{align}
\alpha(\mathbf{x})= \mu({x}) + \kappa \sigma(\mathbf{x}) .
\label{acquisition function}
\end{align}
\noindent where $\mu(\mathbf{x})$ represents the mean of the GP model at point $\mathbf{x}$, $\sigma(\mathbf{x})$ denotes the standard deviation, and $\kappa$ controls the exploration-exploitation trade-off.  The parameter $\kappa$ is generally chosen by the practitioner depending on the case study, with the goal of obtaining the optimal solution in the least number of iterations. In this study, we intentionally set a high value for $\kappa$ in the BO acquisition function. This allows us to perform a thorough exploration of the design space, while still maintaining a small exploitation term to guide the search towards optimal areas. When the exploration term becomes significantly larger than the exploitation term, the BO method essentially behaves like a DoE-based approach with the added advantage of being guided towards the optimal regions. This approach can be considered as an optimisation-guided DoE, combining the strengths of both methodologies to efficiently explore the search space.

%%%%%%%%%%%%%%%%%%%%%%%%%%%%%%%%%%%%%%%%%%%%%%%%%%%%%%%%%%
\subsection{Optimisation algorithm} \label{algorithm}

\begin{figure*}
    \centering
\includegraphics[width=1\linewidth]{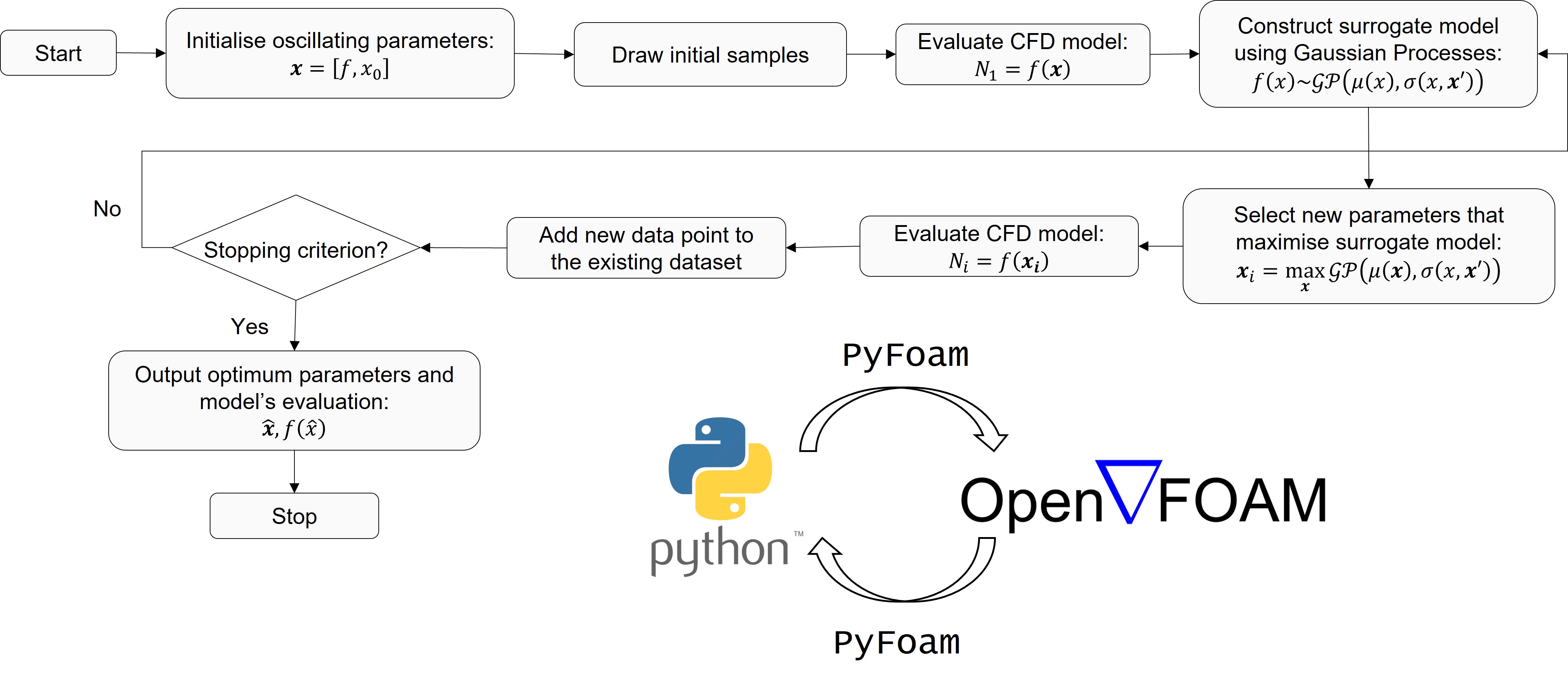}\\
\caption{Flowchart demonstrating the BO-CFD framework.}
    \label{fig:flowchart}
\end{figure*}

\begin{algorithm}
\caption{Exploration-dominant Bayesian Optimisation}\label{bo}
\begin{algorithmic}[1]
    \REQUIRE Objective function $f(\mathbf{x})$, acquisition function $\alpha(\mathbf{x})$,exploration parameter $\kappa$, initial sample points $\mathcal{D}_0 = \{(\mathbf{x}_i, y_i) \}_{i=1}^{n_0}$, number of iterations $n$
    \STATE Initialize dataset $\mathcal{D} \leftarrow \mathcal{D}_0$
    \WHILE{True}
        \STATE Train GP with dataset $\mathcal{D}$
        \STATE \texttt{Select the next point to sample}
        \STATE $\mathbf{x}_{n+1} \leftarrow \arg\max_{\mathbf{x}} \mu(\mathbf{x}) + \kappa \sigma(\mathbf{x})$
        \STATE \texttt{Evaluate the objective function at the selected point}
        \STATE $y_{n+1} \leftarrow f(\mathbf{x}_{n+1})$

        \STATE \texttt{Update the dataset with the new observation}
        \STATE $\mathcal{D} \leftarrow \mathcal{D} \cup \{(\mathbf{x}_{n+1}, y_{n+1})\}$
    \ENDWHILE
    \STATE \texttt{Return the final dataset}
    \RETURN $\mathcal{D}$

\end{algorithmic}
\end{algorithm}

\noindent Algorithm \ref{bo} was implemented in Python 3.10.4, and the Gaussian process (GP) at each iteration was trained using the GPJax library \cite{Pinder2022}. Matern 5/2 \cite{rasmussen2005gpml} is used for constructing the kernel in the GP as it has been evidenced to achieve faster convergence \cite{Morita2022}.
At each iteration, the dataset was normalised to ensure each variable has a mean of 0 and a standard deviation of 1. 
After preliminary experiments, we set the exploration parameter $\kappa=5$, which represents a good trade-off between exploration and optimisation. In the majority of BO cases, a computational `budget' is used as a termination criterion. In this case, we demand the termination of the algorithm when a satisfactory number of samples, or the objective function (corresponding to the equivalent number of tanks-in-series, as will be discussed below), does not change significantly. Each function evaluation consists of an OpenFOAM simulation, which was integrated with Python using the PyFOAM library. The flowchart on the optimisation framework is shown in Figure \ref{fig:flowchart}, $i$ in the subscript indicates iterations, and $\hat{x}$ and $f (\hat{x})$ represent the optimal decision variables and objective function value, respectively, upon termination.

%%%%%%%%%%%%%%%%%%%%%%%%%%%%%%%%%%%%%%%%%%%%%%%%%%%%%%%%%%
\subsection{CFD modelling} \label{numericalmethods}
\noindent To perform an evaluation of a given reactor mesh with a set of operating conditions, a simulation is performed with the open-source code OpenFOAM 1906 version using the finite volume method. A 3D structured mesh for this geometry with the inflation layers close to the walls is generated using a custom mesh generation scheme in Python using the \texttt{classy\_blocks} library, available at \url{https://github.com/OptiMaL-PSE-Lab/pulsed-reactor-optimization/}. Generated mesh for the cell count  167,040 is shown in Figure \ref{fig:meshed_image}b. 

An impulse tracer is injected as a scalar field of concentration $s$ at the inlet of the reactor for the time duration until $t=0.15 \ s$. The laminar solver was selected for this study with water (density: 998.2 kg/m\textsuperscript{3} and viscosity: 0.0010 Pa.s) as the medium. The concentration of the tracer ($s$) is tracked by solving for the convection-diffusion equation through scalarTransportFoam (Eq. \ref{convection-diffusion}):
\begin{align}
\frac{\partial s}{\partial t}+ \nabla \cdot ({\mathbf{v}}s)= D\nabla^2 s,
\label{convection-diffusion}
\end{align}
where $\mathbf{v}$ denotes the flow velocity and $D=1\times 10^{-10}~{\rm m}^2/{\rm s}$ is the diffusion coefficient; thus, the Peclet number $Pe=v_{\rm steady} D_t/D \gg 1$, and the flow is convection- rather than diffusion-dominated. 
Lastly, we normalise $s$ by its inlet value $s_0$ such that 
$s \in [0,1]$. 

The pressure-velocity coupled, transient pimpleFOAM solver is used for solving the unsteady momentum equations as time-dependent oscillatory velocities are introduced. This pimpleFoam solver is integrated with the scalarTransportFoam through `Solver function Objects'. The convection flux on the computational cells was calculated using second-order discretization schemes to ensure the numerical accuracy of the solution. The groovyBC boundary condition is used for imposing oscillatory velocity through swak4Foam library \citep{gschaider2013incomplete}. 
This oscillatory velocity along with the steady velocity was initialised at the inlet as Hagen-Poiseuille parabolic velocity profiles to cut down on the coil length needed for the flow development to reduce the computational cost. 

The numerical solution for every time step was considered as converged when the normalised residuals for the equations solved were lower than 10\textsuperscript{-5}. Additionally, we terminate the CFD evaluation of a case by monitoring the tracer concentration at the outlet; this occurs when the tracer concentration drops to a value less than a tolerance ($s<10^{-7})$ for 10 consecutive iterations. This variable early-stopping criterion based on output accelerates the optimisation procedure, unlike other studies where a fixed termination based on certain number of iterations is enforced \citep{Mansour2020b}. 

The output from a simulation returned from PyFOAM (as the solver is integrated with the optimisation algorithm via the PyFOAM Python library) is a set of concentration values and respective times at the outlet of the reactor. This represents the residence time distribution (RTD) of the reactor. To convert this distribution to a single optimisation objective, the distribution is transformed to an equivalent number of tanks-in-series, $N$. This transformation involved converting time $t$ and concentration values at calculated time $s(t)$ to quantities $E(\theta)$ and $\theta$ using equations \ref{rtd-etheta} and \ref{theta}, respectively.  
\begin{align}
    E(\theta)= \tau E(t), \label{rtd-etheta}
\end{align}
\begin{align}
    \theta= \frac{t}{\tau}, \label{theta}
\end{align}
where $E(t)$ and $\tau$ can be written as,
\begin{align}
    E(t)= \frac{s(t)}{\sum_{0}^{\infty} s(t) \Delta t},
\end{align}
\begin{align}
    \tau = \frac{\sum ts(t) \Delta t}{\sum s(t) \Delta t},
\end{align}
We then fit the tank-in-series model defined in equation \ref{etheta} with the obtained $E(\theta)$ versus $\theta$ curve to determine the value of $N$. The tank-in-series model is given by \citep{McDonough2019a}: 
\begin{align}
    E(\theta) = \frac{N(N\theta)^{N-1}}{(N-1)!}e^{-N\theta},\label{etheta}
\end{align}

While a least-squares fitting procedure based on the $L_2$-norm could be used to fit the tank-in-series model to the calculated $E(\theta)$ versus $\theta$ curve, we observed non-idealities. Therefore, we defined the objective function $N$ to be the value of $N^*$ that maximises the absolute difference between the maximum predicted value and the maximum value obtained from the simulation results, as shown in equation \ref{n_cost}.
\begin{align}
        N^* = \arg\max_N \; \left|\max\left[{E(\theta)}\right] - \max\left[\frac{N(N\theta)^{N-1}}{(N-1)!}e^{-N\theta}\right]\right|. \label{n_cost}
\end{align}

Note, the current modelling approach can be easily extended to problems with fluids of differing physical properties modelled as multiphase flows or with reactions where chemical kinetics can be coupled to CFD simulations.
%%%%%%%%%%%%%%%%%%%%%%%%%%%%%%%%%%%%%%%%%%%%%%%%%%%%%%%%%%%%
\subsection{Characterisation of mixing flow}
\noindent Although the quantification of plug flow performance through a single value $N$ is useful for the optimisation study, it does not  capture well the intricate mixing features associated with axial and radial flows. To gain a more comprehensive understanding of the mixing characteristics as the decision variables vary, we draw inspiration from the helical baffled version of the Oscillatory Baffled Reactor (OBR), which can achieve plug flow over a considerable range of oscillating parameters due to the additional development of swirling motion. Swirling, quantified through the swirl number $S_n$, influences the redirection of flow in the tangential direction and can either enhance or impede axial dispersion.
Similarly, the strength of vortices formed in the coil cross-section, which affects radial mixing, can be  analogously quantified by the radial number $r_n$. These mixing characteristics are expressed by $S_n$ and $r_n$, given by equations \ref{eqn6} and \ref{eqn7}, respectively.

\begin{align}
	S_n &= \frac{\int{\xi^{\dprime}dA}}{R \int{\zeta^{\dprime}dA}},\label{eqn6}\\
	r_n &= \frac{\int{\eta^{\dprime}dA}}{\int{\zeta^{\dprime}dA}}, \label{eqn7}
 \end{align}
where $dA$ is a differential area element in the pipe cross-section, $R=D_t/2$ is the pipe radius, while $\xi^{\dprime}$, $\eta^{\dprime}$, and $\zeta^{\dprime}$ are given by
\begin{align}
    \xi^{\dprime} &= v_yv_{\theta}\sqrt{x^2+z^2},\label{eqn3}\\
    \eta^{\dprime} &= v_yv_r,\label{eqn4}\\
    \zeta^{\dprime} &= v_y^2, \label{eqn5}
\end{align}
$v_\theta$ and $v_r$ denote the tangential and radial velocity components, respectively, expressed by
\begin{align}
	v_{\theta} &= \frac{({xv_z}-{zv_x})}{\sqrt{x^2+z^2}},\label{eqn1}\\
	v_{r} &= \frac{({xv_z}+{zv_x})}{\sqrt{x^2+z^2}}.\label{eqn2}
\end{align}
Here $v_x$, $v_y$ and $v_z$ are velocities in $x$, $y$ and $z$ directions. 
The quantities $S_n$ and $r_n$ are obtained by evaluating the constituent variables of  equations (\ref{eqn6})-(\ref{eqn2}) in the $x-z$ plane
(with origin at (0, 0, 0.005) and normal vector of (0, 1, 0)). In the aforementioned equations, $v_y$ represents the axial velocity since its vector is normal to the x-z plane. The evaluation is done using the `swak4Foam’ library in OpenFOAM which allows for the definition of custom ‘fieldFunctionObjects’.

The introduction of $S_n$ and $r_n$ in this study serves two primary purposes. Firstly, by associating $S_n$ and $r_n$ with the decision variables ($x_0$, $f$) through the dimensionless parameters $St$ and $Re_o$, we obtain an informed representation of how oscillation parameters influence flow mixing characteristics and, consequently, the objective function $N$. Secondly, these numbers provide a valuable means of interpreting the flow features under optimal conditions, leading to a more comprehensive understanding of the mixing behaviour within the system.

%%%%%%%%%%%%%%%%%%%%%%%%%%%%%%%%%%%%%%%%%%%%%%%%%%%%%%%%%%
%%%%%%%%%%%%%%%%%%%%%%%%%%%%%%%%%%%%%%%%%%%%%%%%%%%%%%%%%%

\section{Results and discussion}
\subsection{Validation of the computational model}
\begin{figure}
    \centering
        \includegraphics[width=\columnwidth]{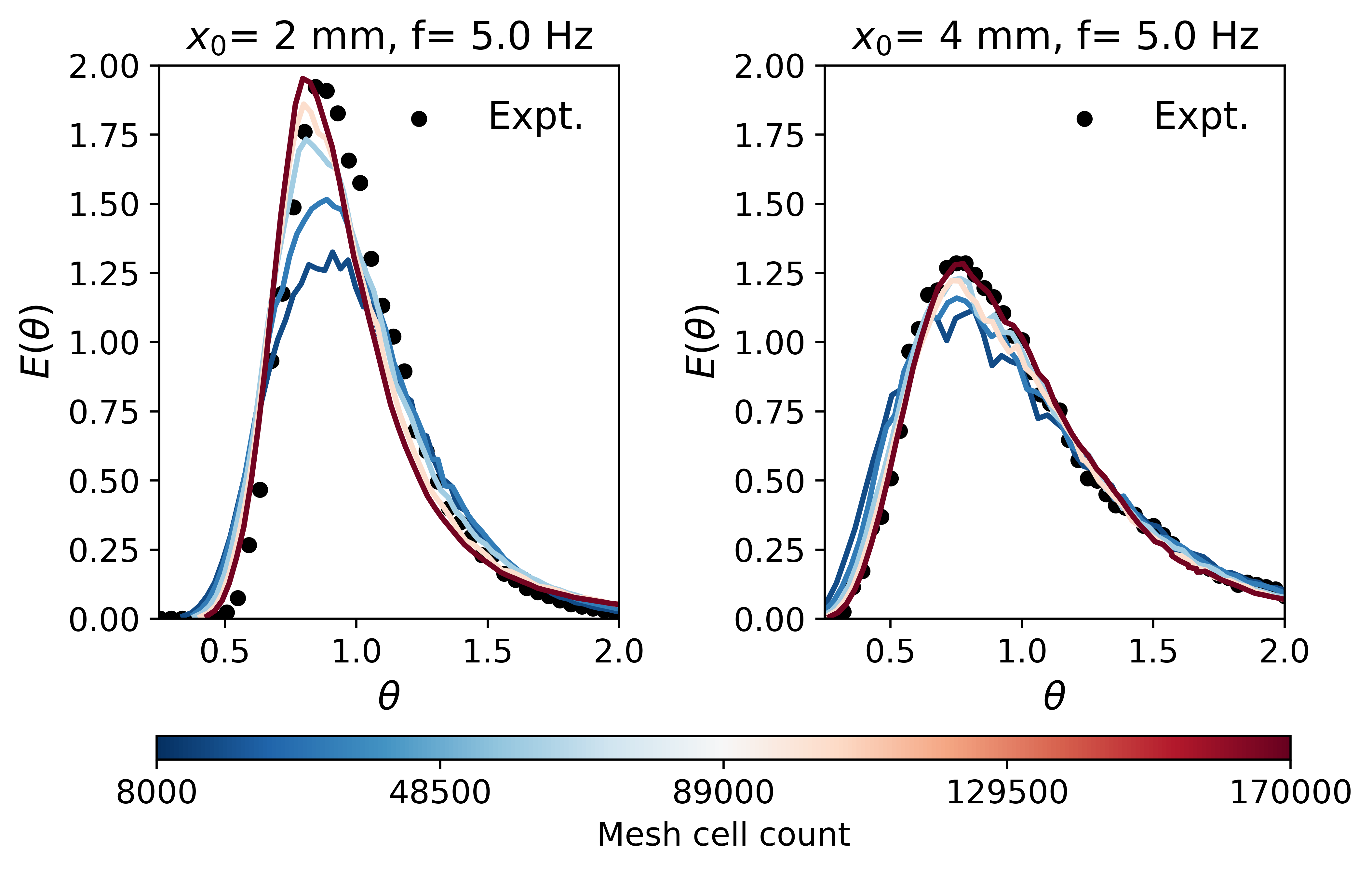}
    \caption{Comparison of dimensionless residence time distribution (RTD) for the computational model with the experimental data for varying cell count.}
    \label{fig:validation}
\end{figure}

\noindent A mesh independence study is conducted for a cell count varying from 8,800 to 167,040 for two different conditions of $x_o= 2.0 \ mm $ and $f= 5.0 \ Hz$, and $x_o= 4.0 \ mm $ and $f= 5.0 \ Hz$ at $Re=50$ (based on the experimental data reported by McDonough et al. \cite{McDonough2019a}). Tracer concentration values were obtained at the outlet of the coil over time and converted to dimensionless residence time distribution (RTD) according to the method described in Section 2.4. The resulting RTD from the computational model is compared with the experimental data and is shown in Figure \ref{fig:validation}. The cell count of 167,040 results in the RTD curve closely matching the experimental values and beyond this value, no further improvement was observed. Therefore, the mesh-independent solution was considered to be obtained and the setup with 167,040 cells was used with the optimisation framework in the rest of the study. 

\begin{figure}
    \includegraphics[width=\columnwidth]{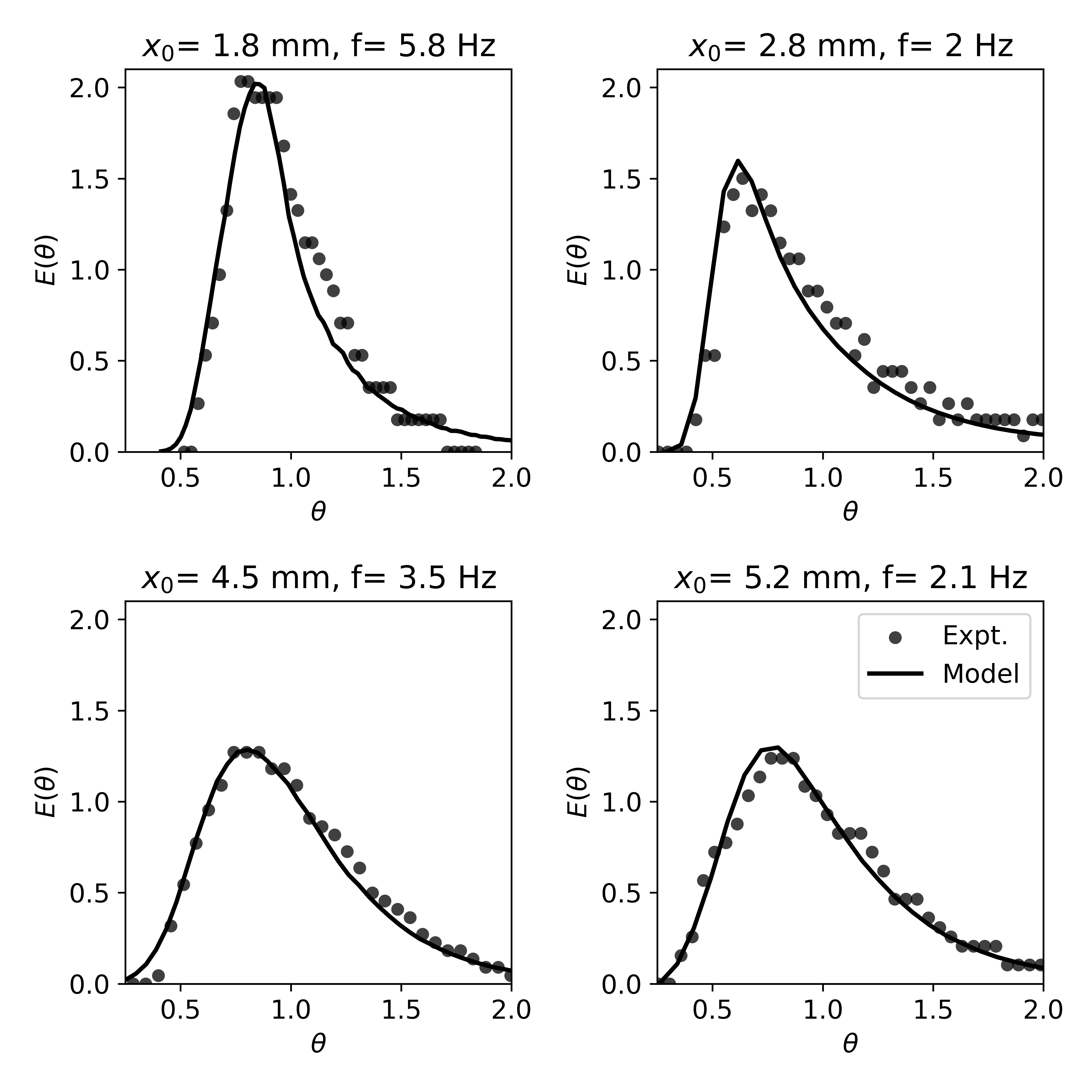}
    \caption{Comparison of dimensionless residence time distribution (RTD) for the experimental data with the numerical model at $Re= 50$, and varied amplitude ($x_o$) and frequency ($f$) conditions. }
    \label{exp_validation}
\end{figure}

Further simulations were conducted for various conditions of $x_o$ and $f$ at $Re=50$ and also compared against experimental data obtained using the same method reported by McDonough et al. \cite{McDonough2019a}. The RTDs were measured by injecting a 0.1 M KCl aqueous tracer solution into the coiled tube and measuring the conductivity over time at the outlet. The net flow of deionized water, oscillations, and tracer injection were controlled using three separate OEM syringe pumps (C3000, TriContinent) that were hydraulically linked to the reactor via PTFE tubing (see McDonough et al. \cite{McDonough2019a} for further clarity).
Figure \ref{exp_validation} demonstrates how predicted values from the CFD model match reasonably well with the experimentally obtained data points for a range of simulated conditions. This gives us sufficient confidence to couple the current setup of the CFD model with the optimisation framework as black-box queries for function evaluations. 

%%%%%%%%%%%%%%%%%%%%%%%%%%%%%%%%%%%%%%%%%%%%%%%%%%%%%%%%%%
\subsection{Exploring decision variables and objective function}
\begin{figure}
    \centering
        \includegraphics[width=\columnwidth]{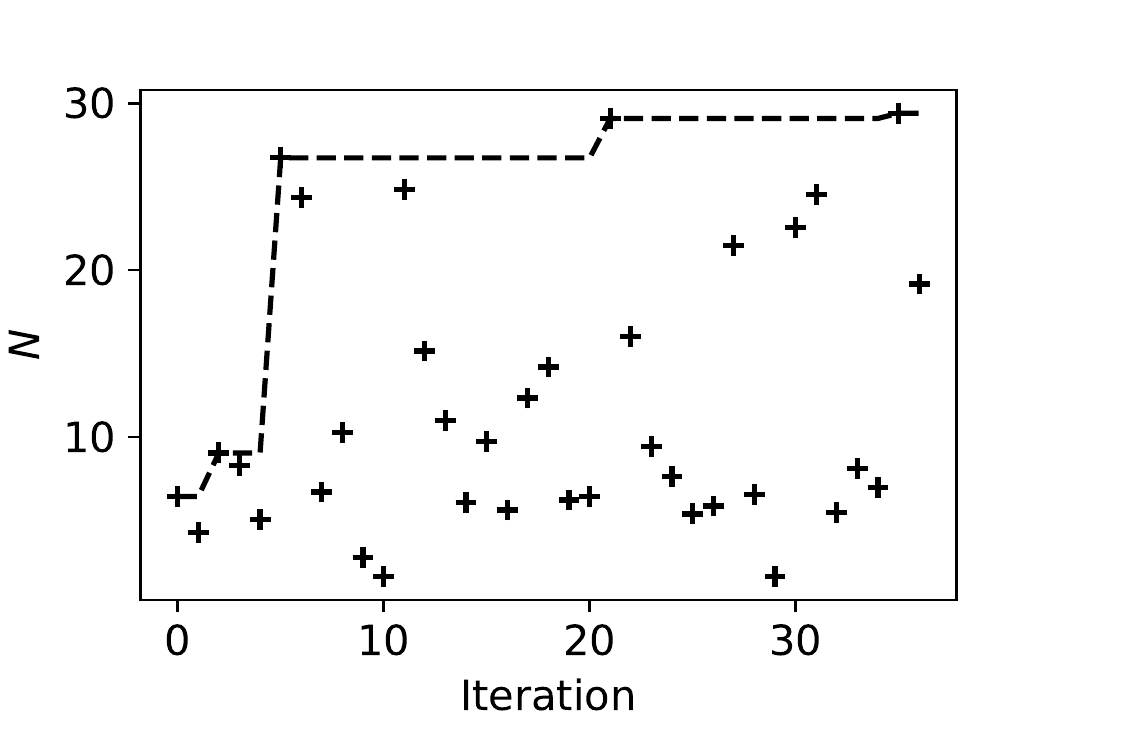}
    \caption{Variation of the objective function, the number of equivalent tank-in-series, $N$, with the evaluations in the design space (`iterations' in the abscissa).} %Optimisation results in the objective function ($N$) space with respect to the iterations.}
    \label{fig:saturation}
\end{figure}

\begin{figure}
    \begin{subfigure}[b]{0.48\textwidth}
         \includegraphics[width=0.95\textwidth]{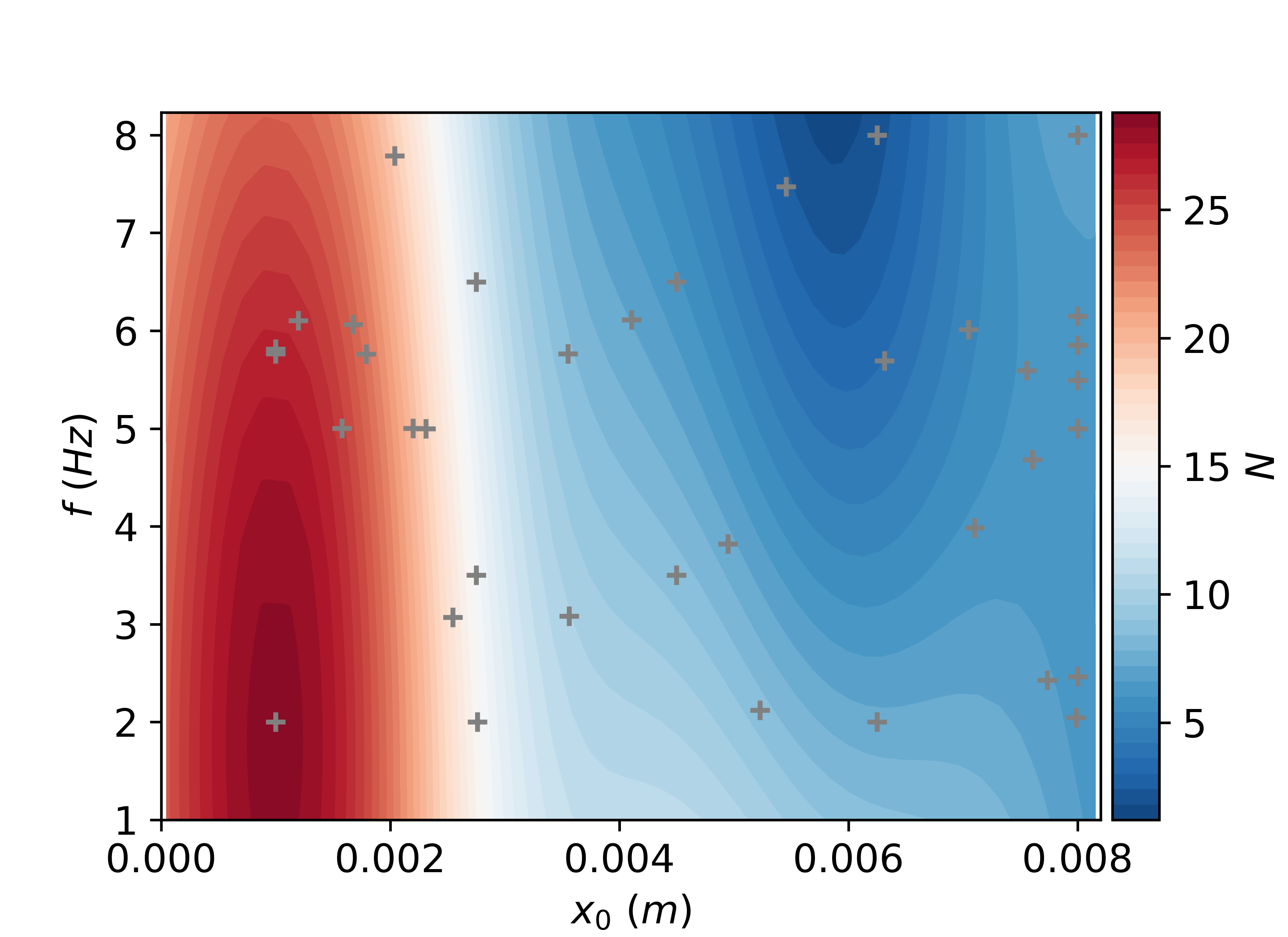}
         \caption{}
         \label{contour_basic_1}
     \end{subfigure}
    \begin{subfigure}[b]{0.48\textwidth}
         \includegraphics[width=0.95\textwidth]{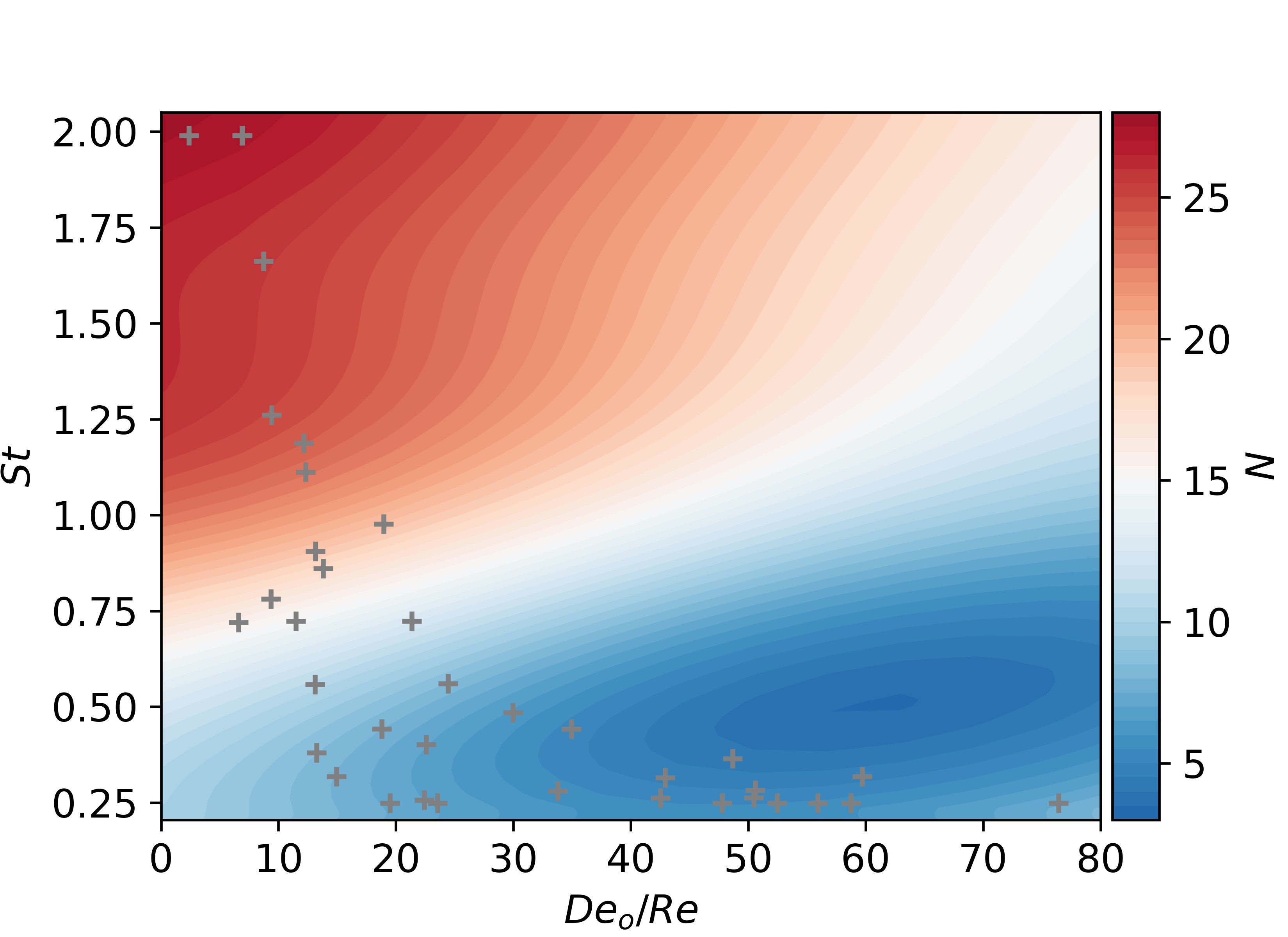}
        \caption{}
        \label{contour_basic_2}
    \end{subfigure}
    \caption{Contour map depicting relationship between plug flow performance ($N$) and (a) amplitude $(x_o)$ and frequency $(f)$, and (b) dimensionless numbers: Strouhal number $(St)$, and ratio of oscillatory Dean number and Reynolds number $(De_o/Re)$.}
    \label{fig:optimal-map}
\end{figure}

\noindent The variation of the objective function $N$ with the iterations is shown in Figure \ref{fig:saturation}. Due to high levels of exploration, the optimiser actively performs function evaluations throughout the design space as it looks to maximise plug flow performance according to the framework developed in the current work (see Algorithm \ref{bo} and Figure \ref{fig:flowchart}). Higher values of expected improvement can be observed within the first 20 iterations compared to the later stages, indicating a potential saturation. The optimisation is terminated after the 36$^{\rm th}$ iteration when the computational budget of 90 hours has been exhausted. 
We note that the choice of time budget is arbitrary and the 90-hour value is set purely for illustrative purposes. 
It is possible, therefore, that an increase in $N$ may result from further iterations though we posit that this marginal gain will be outweighed by the associated increase in computational cost. 

A contour map of $N$ in the $(f,x_o)$ design space is shown in Figure \ref{fig:optimal-map}a where the markers indicate the evaluations conducted. It is seen that $N$ is more sensitive to variations in the amplitude $x_o$ than those in the frequency $f$. For $x_o<2$ mm and $f \in [1-8]$ Hz, $N \in [19.2-29.4]$ which corresponds to good plug flow performance (for reference, $N= 10$ is usually regarded as the minimum acceptable level \cite{mcdonough2019coilincoil}) with the most optimal case being associated with $x_o=1$ mm and $f=2$ Hz. However, for $x_o>2.5$  mm, a significant drop in $N$ occurs, falling from 19.2 to just 1.2 where the latter corresponds to $x_o= 6.25$ mm and $f= 8$ Hz; this indicates that high-amplitude and high-frequency inlet flow oscillations result in poor performance for the geometry considered. 

The results shown in Figure \ref{fig:optimal-map}a are recast in terms of $Re$ as well as $St$ and $De_o$. 
This is shown in Figure \ref{fig:optimal-map}b wherein the $(f,x_o)$ space has been mapped onto that in $(St,De_o/Re)$.
The optimal conditions for $N$ in this space correspond to $De_o/Re \in [2-8]$, with $St \in [1-2]$ which is in agreement with the observations made by McDonough {\it et al.} \cite{McDonough2019a}, obtained through 400 experiments for five different coil geometries in comparison to 36 design space evaluations in 90 hours of computational time. Our BO-based framework, therefore, provides an efficient approach that can yield near-optimal conditions in a fraction of the total experimental time budget.
The `optimal' condition, characterised by $St= 1.98$ and $De_o/Re= 2.8$, for which $N=29.4$ is highest, will be further explored in the subsequent section to `discover' the flow features that minimise axial dispersion and maximise radial mixing.
%%%%%%%%%%%%%%%%%%%%%%%%%%%%%%%%%%%%%%%%%%%%%%%%%%%%%%%%%%
\subsection{Influence of decision variables on objective function via mixing characteristics}
\begin{figure*}
    \centering
\includegraphics[width=1\linewidth]{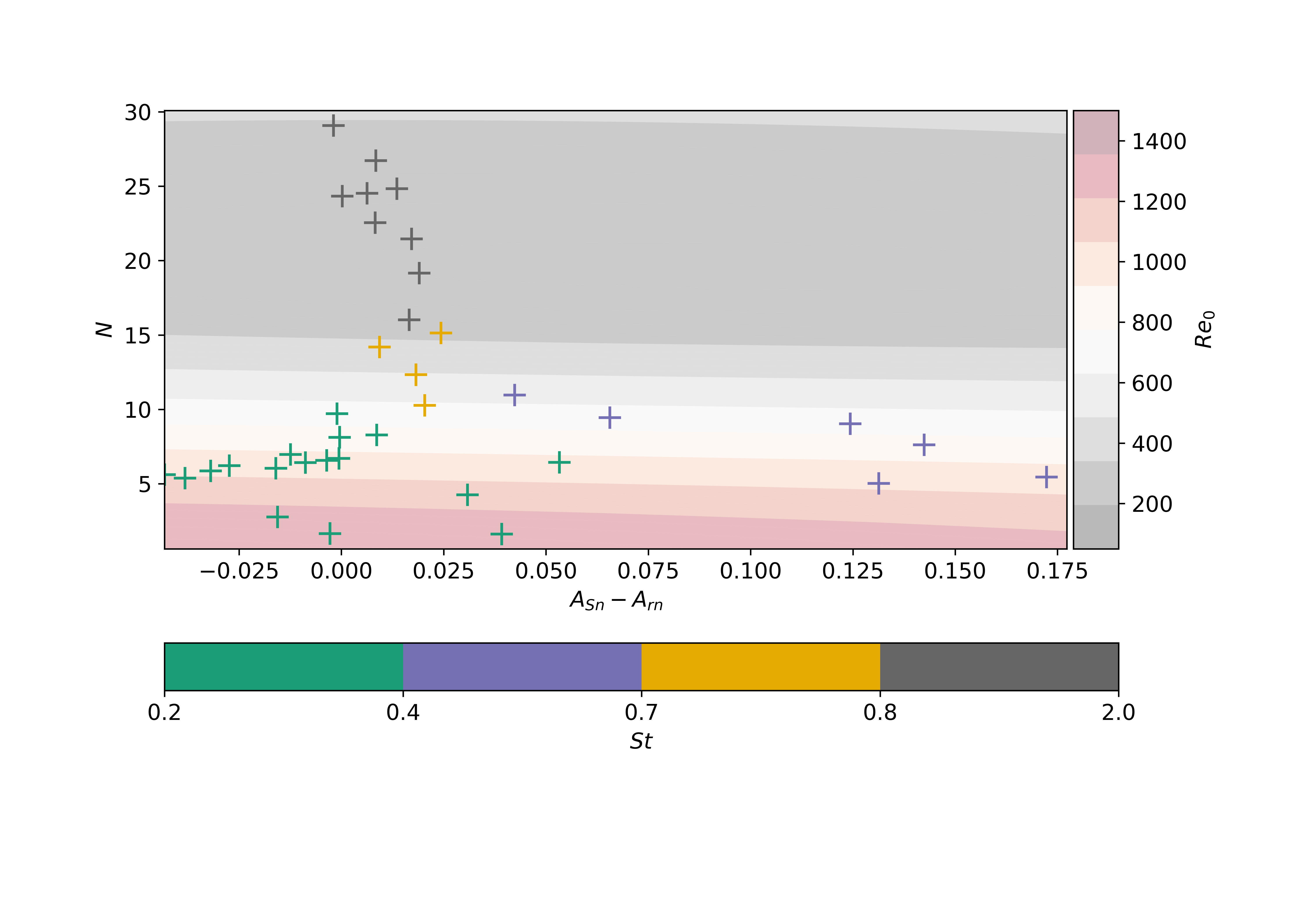}\\
\caption{Contour map showing the relationship of plug flow performance $N$ with the difference in areas of swirl and radial number ($A_{Sn}-A_{rn}$) and oscillatory Reynolds number ($Re_0$). Function evaluation points are grouped according to the Strouhal number ($St$).}
    \label{fig:regime_map}
\end{figure*}

\begin{figure}
    \includegraphics[width=\columnwidth]{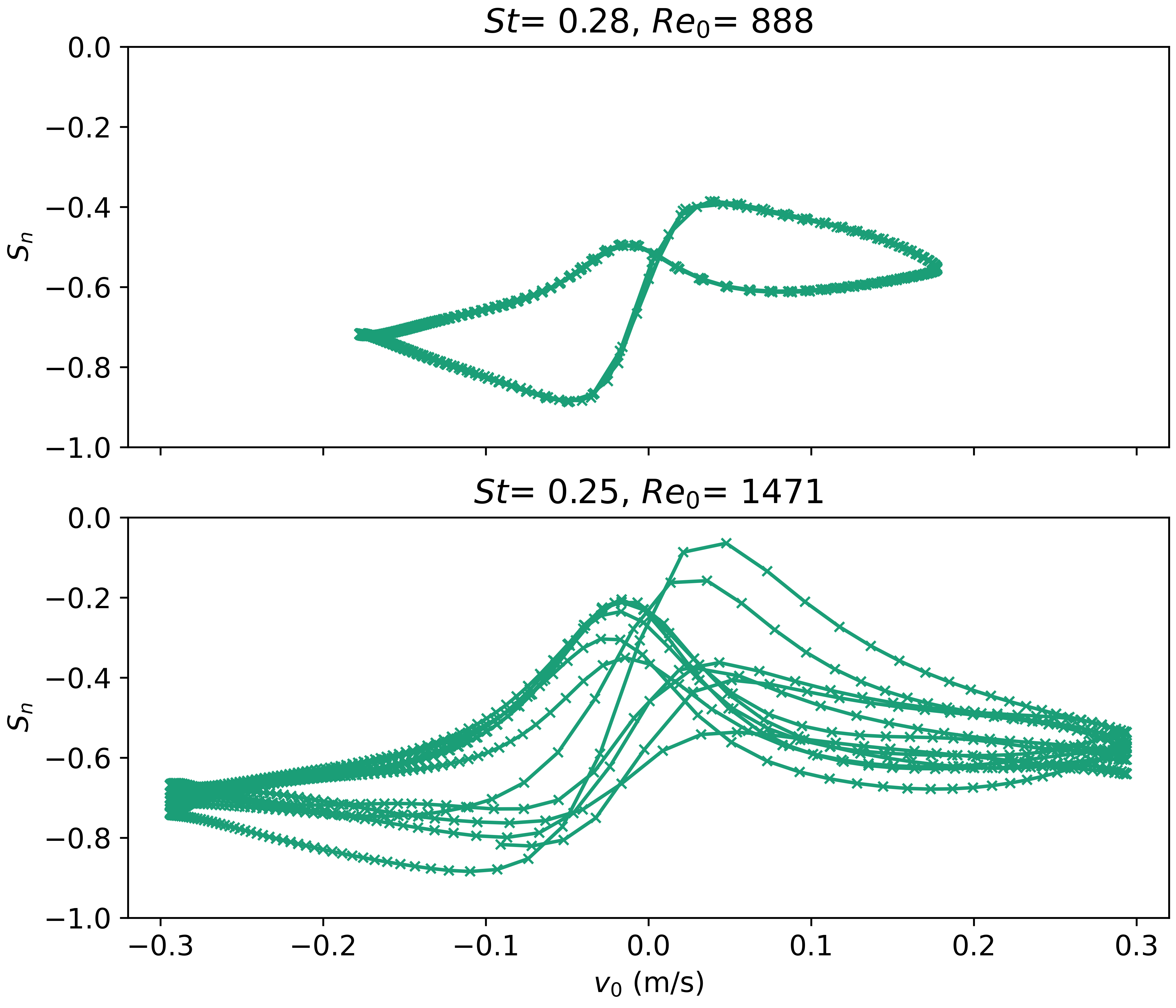}
    \caption{Swirl number $S_n$ with the oscillatory velocity $v_o$ for periodic and non-periodic condition.}
    \label{fig:non_periodicity}
\end{figure}

\noindent To comprehend the relationship between the decision variables $x_o$ and $f$, which are represented as dimensionless quantities $St$ and $Re_o$, and their impact on a specific objective function $N$, we present a graphical representation in Figure \ref{fig:regime_map}. This plot demonstrates the variations in $N$ as a function of $A_{Sn}-A_{rn}$, accompanied by the parametric changes in $St$ and $Re_o$. The values of $A_{Sn}$ and $A_{rn}$ are obtained by calculating the areas under the $S_n$ and $r_n$ orbital plots, respectively (refer to Figure \ref{fig:non_periodicity} for an example of the $S_n$ orbital plot; similar plots can be obtained for $r_n$), and their difference provides a measure of the relative dominance of swirling over cross-sectional vortical flows in the coiled pipe for varying $St$ and $Re_o$.
\begin{figure*}
    \centering
\includegraphics[width=1\linewidth]{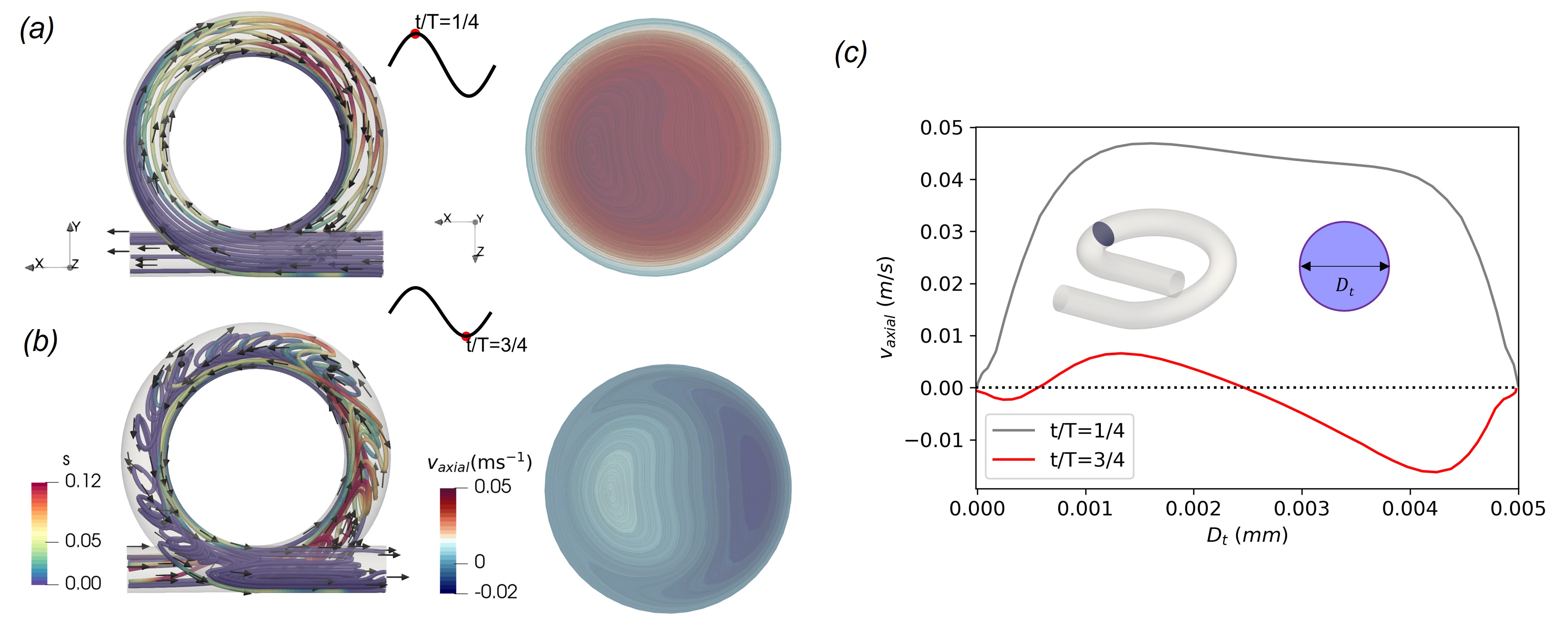}\\
\caption{(a) Velocity streamlines coloured with tracer concentration and the $x-z$ cross-section plane coloured by the streamwise velocity at time intervals  $t/T=1/4$, (b) $t/T=3/4$, and (c) streamwise velocity ($v_{axial}$) across the diameter of the cross-section at $x-z$ plane.}
    \label{fig:streamlines-optimal}
\end{figure*}
\begin{figure}
    \includegraphics[width=\columnwidth]{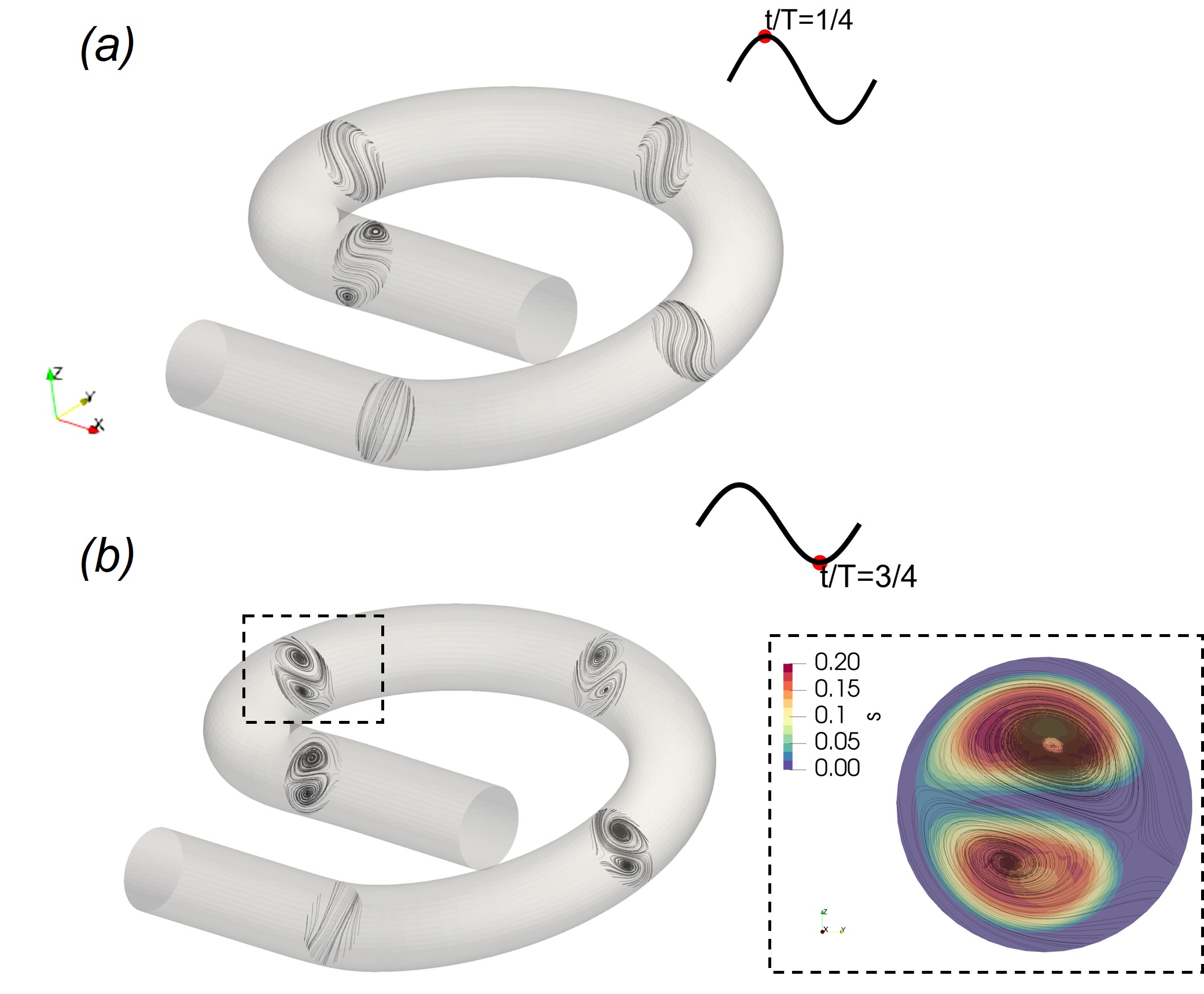}
    \caption{Secondary flow streamlines at various cross-sections of the coil for $t/T= 1/4$ and $t/T=3/4$.}
    \label{fig:vreverse-optimal}
\end{figure}
% A plot of the objective function $N$ as a function of $A_{sn}-A_{rn}$ is depicted in Figure \ref{fig:regime_map}; also shown in this plot is the parametric variation of $St$ and $Re_o$. $A_{sn}$ and $A_{rn}$ are obtained as area under the $S_n$ and $r_n$ curves (refer to appendix for details), and their difference provides a measure of the relative dominance of swirling over cross-sectional vortical flows in the coiled pipe for varying $St$ and $Re_o$. 
Inspired by the work of Sobey \cite{Sobey1983} who showed that the oscillatory flows in obstructed channels exhibit distinct regimes based on the $St$ number, we partitioned our $St$ range into four categories. 

For $St= 0.8-2$, and the lowest $Re_o$ examined, the flow is slightly swirl-dominated, characterised by small $A_{Sn}-A_{rn}$ differences, and is $N$ is approximately in the range of 15-30. 
Among these conditions, the 'optimal' state with $St=1.98$ and $Re_o= 63$ stands out, as it yields the highest $N$ value of 29.8, and we will discuss features pertaining to this condition and the wider $St \in (0.8-2)$.
The slight swirl dominance in this state is supported by the occurrence of swirling streamlines solely during the deceleration phase of the oscillation cycle, specifically shown at $t/T=3/4$ in Figure \ref{fig:streamlines-optimal}a. On the other hand, during the acceleration phase at $t/T=1/4$ in Figure \ref{fig:streamlines-optimal}b, the flow becomes `streamlined'. Consequently, the combination of these two phases results in a minor level of swirling intensity. 

The development of swirling streamlines during the deceleration phase can be attributed to  the differences in the axial component of the flow velocity, $v_{axial}$, around the inner and outer sides of the coil, which is depicted as contours in the $x-z$ plane in Figure \ref{fig:streamlines-optimal}a and b and also plotted as the radial variation of $v_{axial}$ in the same cross-section in Figure \ref{fig:streamlines-optimal}c. 
Notably, $v_{axial}$ assumes small values at the outer coil wall, decreases close to zero, and exhibits a reverse flow along the inner wall. 
In contrast, a uniform distribution of $v_{axial}$ around 0.46 is observed in the cross-section, except at the walls where the no-slip condition is imposed, resulting in a streamlined forward flow without any swirling streamlines.
Nevertheless, the occurrence of minor swirling flow exclusively during the deceleration phase redirects the tracer in the reverse tangential direction, effectively limiting the axial dispersion of the tracer and if not for this swirling, then the tracer would be quickly advected in the forward direction due to the uniform distribution of $v_{axial}$ discussed above. 

\begin{figure}
    \begin{subfigure}[b]{0.48\textwidth}
         \includegraphics[width=0.95\textwidth]{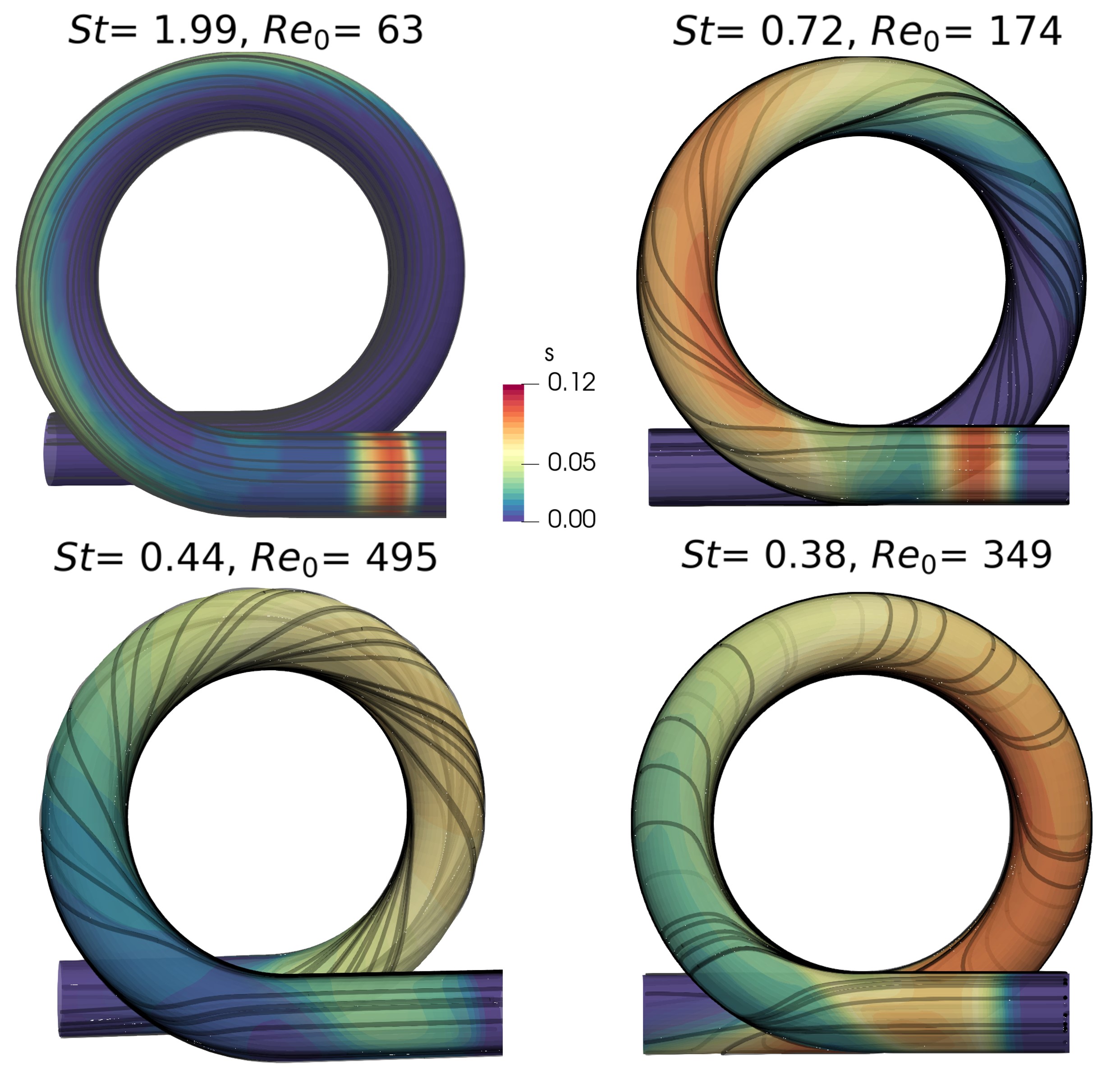}
        \caption{}
        \label{fig:tracer_surface}
    \end{subfigure}
    \begin{subfigure}[b]{0.48\textwidth}
         \includegraphics[width=0.95\textwidth]{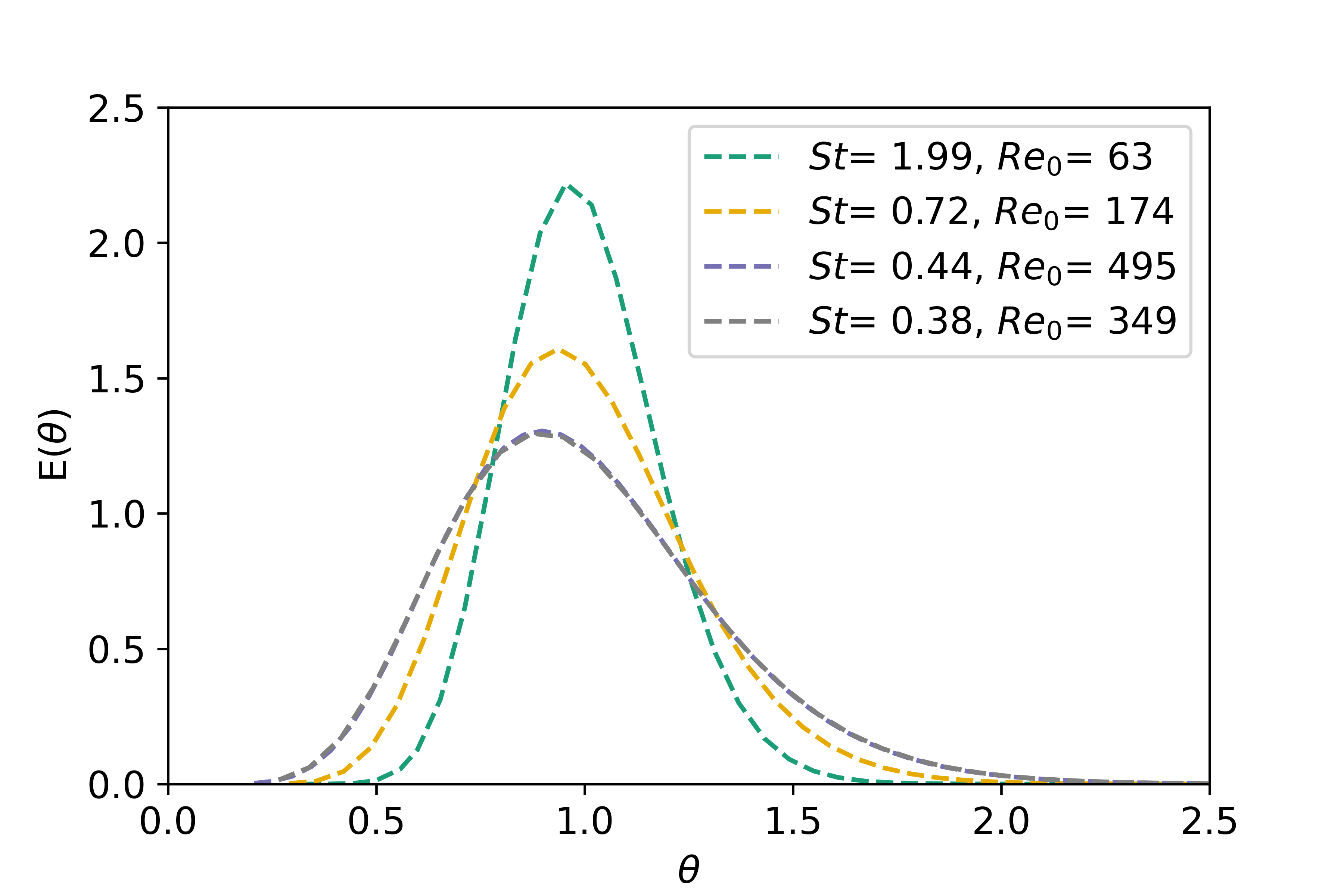}
         \caption{}
         \label{fig:RTD}
     \end{subfigure}
    \caption{(a) Coil surface coloured by tracer concentration and surface streamlines at the interval $t/T=1.0$  and (b) residence time distribution (RTD) for the selected evaluation points based on the $St$ group.}
    \label{fig:rtd}
\end{figure}

The value of $A_{rn}$ can be understood in terms of the secondary flow across the coil cross-section, which contributes to the radial mixing of the tracer.
It is important to note that despite the smaller values of $A_{rn}$ in comparison to $A_{Sw}$, the formation of Dean-type vortices is observed for this case. It is illustrated by secondary velocity streamlines at various coil cross-sections, specifically during the deceleration phase at $t/T=3/4$ in Figure \ref{fig:vreverse-optimal}b, while no Dean vortices are formed during the acceleration phase, as depicted in Figure \ref{fig:vreverse-optimal}a. 
Consequently, it is reasonable to expect that the redirected tracer, caused by the swirling flow, undergoes radial mixing facilitated by the presence of Dean vortices, the radially well-mixed tracer is then advected during the forward phase of the oscillatory cycle. 
As a result of this synergistic effect, a favourable combination of controlled axial mixing and enhanced radial mixing is achieved, leading to an optimal value of $N$ for $St=1.98$ and $Re_o= 63$. Reduced dispersion of tracer is also shown in Figure \ref{fig:rtd}a and narrow variance for RTD in Figure \ref{fig:rtd}b. 

It is important to emphasise that while enhanced radial mixing contributes to the overall performance, the primary impact of reduced axial mixing has a slightly greater influence, resulting in the observed 'good' plug flow performance within the discussed range of $St \in [0.8-2.0]$.
For $St \in [0.7- 0.8]$ and increasing $Re_o$, $N$ drops to the range of 10-15 whilst the flow is slightly more swirl-dominated than for the larger $St$ values. 
This can potentially be attributed to the development of swirling flow, which occurs not only during the deceleration phase but also during the acceleration phase of the oscillatory cycle. The presence of swirling flow streamlines on the coil surface at $t/T=1.0$ supports this observation as depicted in Figure \ref{fig:rtd}a for a condition picked randomly within this $St$ range, specifically $St= 0.72$ and $Re_o= 174$. Moreover, the intensity of swirling flow is higher for both phases due to the larger $Re_o$ values. As a result, this leads to an increased axial dispersion in both forward and backward directions along the length of the coil. This is evident from the wider residence time distribution (RTD) curve for the same randomly picked condition within this range, as depicted in Figure \ref{fig:rtd}b.

We have determined that the $N= 10$ threshold, which should be exceeded for acceptable flow performance, is reached for $St=0.7$. In the ranges $St=0.4-0.7$ and $St < 0.4$, and moderate to large $Re_o$, the dynamics are dominated by swirl and radial flows, respectively. 
In the presence of swirl dominance, similar to the previous $St$ group, there is an increased axial dispersion in both directions, albeit with a more pronounced effect, as depicted in Figure \ref{fig:rtd}a for a randomly picked condition in this range at $St= 0.44$ and $Re_o= 495$. 
On the other hand, with radial dominance, the high-intensity swirling flow is redirected in the radial direction due to growing inertia. This redirection results in a reduced swirling flow angle, as evident from the streamlines in Figure \ref{fig:rtd}a for $St= 0.38$ and $Re_o= 349$. Nonetheless, for both of these groups, the residence time distribution (RTD) exhibits a wider and comparable distribution for the same randomly selected conditions, as shown in Figure \ref{fig:rtd}, and a larger displacement of tracer concentration is observed, as illustrated in Figure \ref{fig:rtd}a. 
As a result, the majority of the emergent $N$ values are below the designated threshold of $N=10$.

Furthermore, for $Re_o> 1100$, due to the accumulating or growing inertia, the flow transitions to a `chaotic-like' state through period-doubling bifurcations, and the associated $N=1.7$ is the lowest calculated in the present study. 
In Roberts and Mackley \cite{Roberts1996}, the transition to this chaotic-like regime occurred at approximately $Re_o> 200$ for a 25mm diameter column containing orifice baffles. This was due to the baffles inducing shear instabilities that enhance the onset of chaotic flows as there exists a dynamic breakup and interaction of multiple vortices \cite{zheng2007development}. In McDonough et al. \cite{Mcdonough2017}, the chaotic regime occurred at $Re_o> 503$ for a 5 mm diameter tube containing a helical coil. The helical flow clearly delays the onset of the chaotic-like state. 
The non-periodic nature of the flow is illustrated clearly in 
Figure \ref{fig:non_periodicity}, which compares $S_n$ over five oscillation cycles for $(St,Re_o)$ combinations of 
$(0.28,888)$ and $(0.25,1471)$, shows that the flows associated with the first and second set of parameters are periodic and non-periodic with $N \approx 7$ and $N \approx 1$, respectively. 

Therefore, our findings indicate that a small value of swirl strength ($A_{Sn}-A_{rn}= 0.0004- 0.019$) at $Re_o= 60-300$ and $St= 0.8-2.0$, associated with the periodic flow regime, results in a `good' plug flow performance. This is reminiscent of the findings of McDonough et al. \cite{Mcdonough2017} where similar observations were made for a helical baffled reactor.
Moreover, it is noteworthy that the conditions of optimality obtained through the BO framework fall within this range, specifically with $St= 1.98$ and $Re_o= 63$. These results highlight the significance of the identified range in achieving desired plug flow conditions for a coiled reactor.
%%%%%%%%%%%%%%%%%%%%%%%%%%%%%%%%%%%%%%%%%%%%%%%%%%%%%%%%%%
%%%%%%%%%%%%%%%%%%%%%%%%%%%%%%%%%%%%%%%%%%%%%%%%%%%%%%%%%%
\section{Conclusions}
\noindent In this study, we propose and apply a derivative-free data-driven Bayesian optimisation (BO) approach to maximise the plug flow performance as an objective function (in terms of the number of well-mixed tanks-in-series, $N$) of a coiled tube reactor by exploring the effects of decision variables, oscillation frequency ($f$) and oscillation amplitude ($x_0$) at fixed Re= 50. The effectiveness of the framework which is an integration between the flow solver and a BO optimiser is demonstrated through the achievement of a near-optimal solution after a relatively modest number of iterations. Hence, this study has resulted in an automated, open-source, cost-efficient method to optimise plug flow performance. 

Additional investigation has shed light on the relationship between the decision variables, mixing characteristics, and the objective function. It was seen that the periodic flow and slight dominance of swirling intensity for $Re_o<300$ and $St= 0.8-2.0$ corresponds to an `optimal' region of performance. Otherwise, a very high swirling intensity or radial intensity results in increased axial dispersion in both the forward and backward directions with $N$ dropping to below the minimum acceptable value of $N=10$. 
By comparing the mixing characteristics between the optimal and non-optimal regions, we have used this CFD-BO approach to `discover' the underpinning mixing patterns that are desirable within this coiled tube geometry subjected to oscillatory flow. Of particular importance, the best performance, characterised by $N=29.8$, was attained when $St=1.98$ and $Re_o=63$, which lies within this `optimal' performance region. Under these conditions, a slightly swirl-dominated flow during the deceleration phase exclusively resulted in minor swirling intensity. The swirling flow, along with the presence of Dean-type vortices during the deceleration phase, contributed to the controlled axial mixing and enhanced radial mixing. 

We believe that by leveraging the findings from this study, future oscillatory reactors can be designed with improved mixing characteristics and enhanced plug flow performance. This can lead to more efficient and cost-effective processes, better reaction control, and higher yields in a range of applications. Moreover, the exploration-led methodology employed in this study opens doors to further exploration; enabling us to unlock the potential of additional parameters such as parametric geometries, different fluids and chemical kinetics, enabling data-driven design approaches for novel reactors and transformative applications. Furthermore, this study serves as a solid foundation for further research and development in flow control and optimisation, thus paving the way for innovative approaches and advancements in reactor design for the future.
\\
%%%%%%%%%%%%%%%%%%%%%%%%%%%%%%%%%%%%%%%%%%%%%%%%%%%%%%%%%%
%%%%%%%%%%%%%%%%%%%%%%%%%%%%%%%%%%%%%%%%%%%%%%%%%%%%%%%%%%

\centerline{\bf Acknowledgements}

\noindent This work is supported by the Engineering and Physical Sciences Research Council, United Kingdom, through the EPSRC PREMIERE (EP/T000414/1) Programme Grant. Tom Savage would like to acknowledge the support of the Imperial College President\textquotesingle s scholarship. We also acknowledge the HPC facilities provided by the Research Computing Service (RCS) of Imperial College London.

%\bibliographystyle{model2-names}


\bibliographystyle{IEEEtranN}

\bibliography{references}
\end{sloppypar}
\end{document}